\ifdefined \singleCol
\documentclass[journal,comsoc,12pt,onecolumn,draftclsnofoot]{IEEEtran}
\else
\documentclass[journal]{IEEEtran}
\fi
\usepackage[T1]{fontenc}%
\usepackage[noadjust]{cite}
\usepackage{graphicx}
\graphicspath{{figures/drawings/}{figures/plots/pdf/}}
\usepackage[export]{adjustbox}
\usepackage{subfig}
\usepackage[cmintegrals]{newtxmath}

\usepackage{capt-of}
\usepackage{color,soul}

\begin{document}
	\title{Signal Processing Based Deep Learning for   Blind Symbol Decoding and Modulation Classification}
	\author{Samer~Hanna,~\IEEEmembership{Student~Member,~IEEE,}
		Chris~Dick,~\IEEEmembership{Senior~Member,~IEEE,}
		and Danijela~Cabric,~\IEEEmembership{Fellow,~IEEE}%
		\thanks{Samer Hanna and Danijela Cabric are with the Electrical and Computer Engineering Department, University of California, Los Angeles, CA 90095, USA. Chris Dick is with Nvidia Inc.  and was at Xilinx Inc. San Jose, California, USA  when this work started  	 e-mails: 	\mbox{samerhanna@ucla.edu}, cdick@nvidia.com, danijela@ee.ucla.edu}
		\thanks{This work was supported in part by the CONIX Research Center, one of six centers in JUMP, a Semiconductor Research Corporation (SRC) program sponsored by DARPA.}
	}
	
	\maketitle

	\begin{abstract}
		Blindly decoding  a signal requires estimating its unknown transmit parameters,  compensating for the wireless channel impairments, and identifying the modulation type. While deep learning can solve complex problems, digital signal processing (DSP) is interpretable and can be more computationally efficient. To combine both, we propose the dual path network (DPN). It consists of  a signal path  of DSP operations that recover the signal, and a feature path of neural networks that estimate the unknown transmit parameters.  By interconnecting the  paths over several recovery stages, later stages  benefit from the recovered signals and reuse all the previously extracted features. The proposed design is demonstrated to provide 5\% improvement in modulation classification compared to  alternative designs lacking either feature sharing or access to recovered signals. The estimation results of DPN along with its blind decoding performance are shown to outperform a blind signal processing algorithm for BPSK and QPSK on a simulated dataset. An  over-the-air software-defined-radio capture was used to verify DPN results at high SNRs. DPN design can process variable length inputs and is shown to outperform  relying on fixed length inputs with prediction averaging on longer signals by up to 15\% in modulation classification. 
	\end{abstract}

	\IEEEpeerreviewmaketitle
	
	\section{Introduction}
	In a typical wireless communication system, the transmitter and receiver exchange waveforms following an agreed upon protocol.    However, a prior agreement on waveforms is not always  possible  and heterogeneous radios  might need to communicate without a protocol predefining waveforms. Without a known protocol, a blind receiver  has to reconstruct and  decode an unknown waveform in many applications. Civilian applications include   decoding  unknown signals to enable exchanging messages between heterogeneous radios using dynamic spectrum access.  Military applications of blind decoding include intercepting hostile communications.

	Decoding a blindly received signal is a challenging problem. In addition to all the channel impairments facing protocol based communications,  a blind receiver lacks any knowledge about the transmitted signal. Without   an agreed upon preamble for  synchronization and channel estimation and without knowing the type of modulation,  a blind receiver has only access to a sequence of  IQ samples representing unknown symbols. From this sequence,  the receiver needs to figure out the symbol rate, the  modulation type, the channel impulse response for equalization, and compensate for carrier and timing synchronization errors to recover the transmitted symbols. Additionally, the length of a blindly received signal is not determined apriori. 
	
	While it is easy to  generate signals of different parameters and channel distortions, developing signal processing algorithms that jointly estimate all these parameters for blind decoding is more challenging. Many existing works have leveraged blind signal processing to address blind decoding and modulation classification~\cite{rebeiz_energy-efficient_2014,kazikli_optimal_2019,liu_blind_2019}. However, these works address the estimation problems separately, thus, not benefiting from  joint estimation, and many of them make  assumptions impractical for a blind receiver like assuming synchronization.

	Driven by the availability of training data, some of the estimation problems in blind decoding were separately addressed  using deep learning;  Center frequency and timing offset estimation neural networks were proposed in a non-blind context for QPSK signals~\cite{oshea_learning_2017}. For modulation classification, deep neural networks were  shown to outperform   manually designed features  in~\cite{oshea_over--air_2018}.  However, posing blind decoding as a set of independent deep learning estimations has its drawbacks.  It does not allow the networks to share common features and thus prevents them from performing joint estimation by learning the dependencies between the  parameters present in the signal. It also denies each network benefiting from the other networks' outputs, which can partially recover the signal and reduce distortions. To avoid these limitations, another option is an end-to-end deep learning solution taking the signal as input and outputting the symbols.

	Using a black box end-to-end neural network for blind decoding also has its drawbacks. Unlike modulation classification or estimation, blind decoding needs to be applied  to the entire length of the signal and thus needs to be efficiently scalable. Deep learning can have a high computational complexity compared to  digital signal processing (DSP) and lacks interpretation~\cite{balatsoukas-stimming_deep_2019}. For efficient scaling, estimating the unknown parameters and relying on DSP for compensation can be more computationally efficient than relying on an end-to-end black box neural network. Additionally, interpretable outputs can easily be integrated with  existing DSP techniques. The drawbacks of separate estimations and end-to-end solutions motivate for a model-based deep learning solution~\cite{he_model-driven_2019}, integrating both signal processing and deep learning in a  design combining the benefits of both.
	
	Another common limitation of neural networks used in many existing works addressing similar problems is the fixed input size.  For instance,  in~\cite{oshea_over--air_2018}, it was shown that neural networks using longer signals can lead to improved modulation classification accuracy. However, the proposed networks can only process fixed length  signals and network redesign and retraining is required to change the input length. Blind receivers do not have control over the received signal length and it is not practical to train a separate network for each possible signal length. To use fixed length networks on variable  size inputs, one approach  is to train them on  short signals and divide longer signals into chunks and use averaging as proposed in~\cite{zheng_fusion_2019}. Another approach, which provides more flexibility, is to  design the neural networks that handle variable length signals as proposed in~\cite{rajendran_deep_2018-2,courtat_light_2020}. However, it is not clear which approach yields a better performance.

	In this paper, we propose the Dual Path Network (DPN) for blind decoding and modulation classification. To overcome the limitations of separate estimations and end-to-end networks, the dual path network is designed as two paths; a signal path and a feature path. The signal path consists of linear signal processing operations where  frequency offset, noise, and fading are compensated for sequentially. The feature path consists of neural networks  that process the features extracted from the input and the compensated signals. These features are used to predict interpretable signal processing estimates and filter taps.  Using this architecture, DPN benefits from recovered signals and can combine features from all stages. Once the estimates have been obtained, due to their signal processing interpretation, they can be integrated with  existing signal processing techniques. By leveraging average pooling and recurrent networks to obtain the predictions, DPN  can process variable length inputs.  Our contributions can be summarized as follows
	\begin{itemize}
		\item We propose the Dual Path Network (DPN) architecture for blind signal decoding, which can process inputs of variable lengths. DPN design integrates  signal processing  with neural networks yielding interpretable outputs. The proposed  architecture enables learning from recovered signals and allows features combining.
		\item  We verify the benefit of learning from recovered signals and combining features along the signal path. To do that, we  compare DPN to alternative architectures that lack access to either recovered signals or feature combining and show that DPN outperforms either architecture by up to  5\% in modulation classification. 
		\item   For inference on variable length, we show that using DPN trained and tested with variable lengths provides lower estimation errors and up to 15\%  improvement in modulation classification  compared to   averaging the predictions calculated using the same network trained and tested only using a short fixed input size. 
		\item We  show that DPN provides lower estimation errors  and a higher percentage of correctly decoded signals for BPSK and QPSK on the test dataset compared to an implemented blind signal processing algorithm.
	\end{itemize}

	\section{Related Work}
	In  section, we start by discussing existing signal processing approaches for blind decoding, then we survey the deep learning approaches used for problems that are part of blind decoding. Works that have considered deep learning   demodulators are discussed last.
	\paragraph{DSP approaches for Blind Decoding}
	Blind decoding DSP approaches rely on recovering the symbols then analyzing them to identify the modulation types. In \cite{rebeiz_energy-efficient_2014}, cyclostationary features were used for  carrier frequency offset and symbol rate estimation to recover the symbols. Using the recovered symbols, a decision tree algorithm based on statistical tests  for  blind modulation classification  was proposed.    Assuming  synchronization in an AWGN channel, joint modulation classification and symbol decoding were performed using two approaches based on a Bayesian framework and a minimax framework in~\cite{kazikli_optimal_2019} . 
	Under the assumption of perfect synchronization, in \cite{liu_blind_2019}, an iterative approach  was proposed for joint blind channel estimation, modulation classification, channel coding recognition, and data detection using a different approach for each task. 
	\paragraph{Deep Learning Approaches for Estimation and Modulation Classification}
	Many  deep learning approaches were proposed for problems relevant to blind decoding. Deep learning was proposed for carrier frequency offset  and timing offset estimation and was compared to traditional estimators in \cite{oshea_learning_2017}. 
	In \cite{oshea_convolutional_2016}, deep learning from raw IQ samples was shown to outperform manually designed features in the problem of modulation classification. This result sparked a wide interest in developing novel modulation classification neural network architectures. 
	In~\cite{oshea_over--air_2018}, a residual neural network was proposed and the effects of fading and frequency offsets were studied.  Recurrent neural networks were proposed for modulation classification in~\cite{hong_automatic_2017} and were shown to outperform convolutional networks. A network combining both  CNN and LSTM networks was proposed in~\cite{zhang_automatic_2020-1}. In \cite{vanhoy_hierarchical_2018},  branch convolutional neural networks, consisting of a  hierarchy of neural networks were used to classify 29 modulation types.
	A  network design that  combines shallow and high level features of the input signals was proposed in~\cite{nie_deep_2019}. In~\cite{chen_novel_2020}, a cyclic connected CNN along with a bidirectional RNN were proposed.  Instead of using IQ samples, the signals to be classified were represented as constellation images  in~\cite{peng_modulation_2019,huang_automatic_2019,huang_automatic_2019-1,doan_learning_2020}. Other works have focused on designing lightweight networks that reduce the number of parameters or the inference time~\cite{ramjee_fast_2019,ramjee_ensemble_2020,ke_real-time_2020,icamcnet_2020,zhang_automatic_2020,liao_sequential_2019,courtat_light_2020}. Modulation classification was also considered for multi-carrier signals in~\cite{xu_deep_2019} and for MIMO systems in~\cite{li_deep_2020,jiang_time-frequency_2020}. In~\cite{rajendran_deep_2018-2,wong_distributed_2019}, distributed modulation classification using multiple receivers to capture the same signal was proposed.
	
	Deep learning was shown to be affected by synchronization errors, which motivated the development of  custom signal processing layers. The effects of carrier frequency offsets and sampling offsets on modulation classification were studied in~\cite{hauser_signal_2017}.  It was shown that frequency offsets and sampling rate offsets degrade the classification accuracy with the former  having a more pronounced effect. Driven by this observation, custom neural network layers that attempt to recover the signal before classification were proposed. A spatial transformer network was proposed in~\cite{hanna_asilomar_2017} for timing recovery and was shown to improve accuracy at low oversampling. In~\cite{oshea_radio_2016}, a radio transformer network that can correct frequency offset and adjust the sampling rate of the signal was proposed. Similarly, a learnable distortion correction module for frequency and phase offset was proposed in ~\cite{yashashwi_learnable_2018}.  In \cite{shang_dive_2020},  a U-net network was proposed to reconstruct low SNR signals. In these works,  the custom reconstruction layers were trained to improve the modulation classification results  without the reconstruction  ground truth, and hence are not guaranteed to recover the transmitted signals. Thus, the output of such networks can not be used for blind decoding.
	
	Most of the neural network architectures in the literature were designed only to process signals with a fixed length input. To use a fixed length network on a signal with a different length, the network has to be redesigned and retrained.  To extend a network to longer signals without retraining, in \cite{zheng_fusion_2019}, the authors have proposed methods to combine the predictions applied on small chunks of a signal.  These methods  consist of averaging the predictions. Neural networks that can process variable input size signals were proposed by using LSTM in~\cite{rajendran_deep_2018-2} and by combining convolutional layers and average pooling in~\cite{courtat_light_2020}.  However, it is still not clear which network design yields a higher accuracy on signals of length unseen in training;   a fixed length network with combined predictions or a variable length network. 

	\paragraph{Deep Learning Approaches for non blind Demodulation}
	Several works have proposed using deep learning for demodulation when the transmitter signal parameters are known apriori. In~\cite{zheng_demodnet_2020}, an end-to-end  neural network demodulator was developed  under the assumption of time synchronization between the transmitter and receiver. Several modulation types were supported and a network was trained for each type.
	For WiFi,  neural networks were proposed  to replace several signal processing blocks of a WiFi receiver in~\cite{zhang_deepwiphy_2020}. These blocks  include channel estimation, phase error correction, sampling rate correction, and equalization and rely on the WiFi preambles to perform their tasks. A similar approach was proposed for 5G compliant waveforms  in~\cite{honkala_deeprx_2021}. Deep learning end-to-end communications systems, where both the transmitter and receivers are neural networks, were also proposed. In \cite{dorner_deep_2018} and~\cite{oshea_learning_2016} autoencoders were proposed for SISO communications and in \cite{oshea_deep_2017}  for MIMO. These approaches, however, rely on knowing or designing the transmitted signal and do not directly apply to blind decoding.
	
	The dual path network was first introduced in our prior work~\cite{hanna_combining_2020}, which we extend in this work. In this work, DPN design and  training process was improved as discussed later. A more thorough evaluation was performed to evaluate  the effect of signal reconstruction and  feature sharing among different stages. The blind decoding performance was compared with both blind and Genie signal processing approaches. An over-the-air capture was used to validate our results. DPN was evaluated on variable input lengths and different approaches for variable length training and inference were considered.

	\renewcommand{\b}[1]{\boldsymbol{\mathrm{#1}}}
	\newcommand{\mC}{\mathbb{C}}
	\section{System Model and Problem Formulation}
	We start by  describing the system model and the underlying assumptions of our work. A transmitter sends a vector of  complex symbols $\b{s} \in \mC^{N_s}$ using modulation type $M$ from a set of modulations $\mathcal{M}$. In the most general case, the transmitted signal $x(t)$ is determined by symbols $\b{s}$ and symbol duration $\tau$ through a modulation specific mapping function $\mathcal{G}$  such that
	$x(t) = \mathcal{G}(\b{s},\tau)$.
	For a linear modulation type, each symbol $s_i$ represents a mapping from bits to a  constellation point, and the transmitted signal $x(t)$ is given by 
	$
	x(t) = \sum_{i=1}^{N_s} s_i p(t-i \tau)
	$
	where $p(t)$ is the pulse shaping filter. The signal is upconverted and  transmitted over a multipath fading channel modeled with an impulse response $h(t)$ having a delay spread $\sigma$. At the receiver, the downconverted and sampled signal is  modeled using the vector $\b{y} \in \mC^{N_r}$ with the $k$-th element given by
	\begin{equation}
	y[k]= e^{j 2\pi( f_0 t_k + \phi_0)} \int_{-\infty}^{\infty} x(\gamma) h(t_k -\gamma) d\gamma  + n(t_k)
	\end{equation}
	where $f_0$ is the carrier frequency offset, $\phi_0$ the phase offset, and $n(t)$ is the additive white Gaussian noise process with zero mean and power spectral density $N_0/2$. We assume that the receiver sampling rate is given by $\frac{1}{\tau_{0}}$. The sampling instance of the $k$th sample is given by  $t_k=t_{0} + k \tau_{0}$, where $\tau_{0}\leq \tau$ and $t_{0}$ is the symbol timing offset. 
	The length of the transmitted symbols $N_s$ and the received IQ samples $N_r$ are related using
	$N_r = N_s \left\lceil \frac{\tau}{\tau_0} \right\rceil$. 
	Such signals can  be captured using a narrowband receiver coarsely tuned to the center of a spectrum occupancy or using a wideband receiver followed by coarse band segmentation.  Either way, we assume that the vector $\b{y}$ contains only one signal and that the sampling  satisfies the Nyquist criterion. The  parameters $f_0,\phi_0, h$ are assumed to be constant along the duration of each signal.  %
	
	To avoid dependence on hardware specifications that vary from one radio to the other, we normalize the previous  parameters with respect to the sampling rate. This is done by  using the frequency normalized to the sampling rate $\frac{f_0}{1/\tau_0}$,  the number of samples per symbol $\frac{\tau}{\tau_0}$, and   the normalized timing offset $\frac{t_0}{\tau_0}$.  Without loss of generality,  to simplify the notation, we consider $\tau_0=1$, which is equivalent to assuming a 1Hz sampling rate. Under this assumption, the number of samples per symbol $\tau$ becomes equal to the symbol duration. 
	
	The problem considered is described as follows: Given the vector $\b{y}$, the receiver's objective is to identify the modulation type $M$ among the set of modulations  $\mathcal{M}$ and recover the transmitted symbols~$\b{s}$.

	\section{Dual Path Network (DPN)}
	\begin{figure*}[t!]
		\centerline{\includegraphics[scale=1]{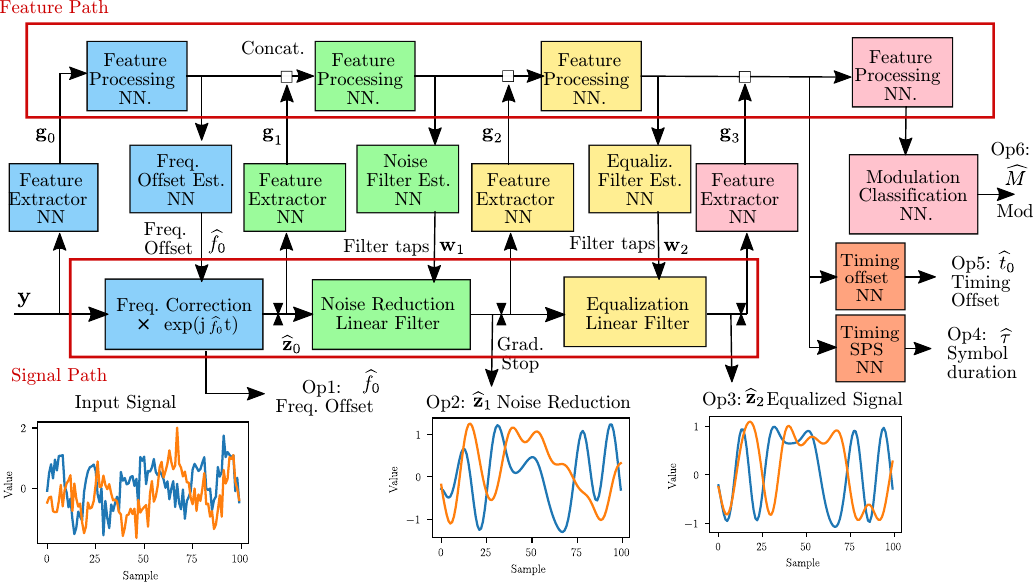}}
		\caption{The Dual Path network consists of feature  path and a  signal path connected using neural networks (NN) for parameter estimation and feature extraction. A slice of an example input signal and expected outputs of the network are plotted. }
		\label{fig:dualPath}
	\end{figure*}
	The main ideas behind the dual path network  design are: (1) the less the distortions, the better the predictions, (2) Estimating jointly is better than separately. If the receiver had access to the transmitted signal, $x(t)$, the considered problem would have been easier. Using this idea, to improve its predictions, DPN attempts to recover the signal $x(t)$. The signal recovery is inspired by existing DSP techniques, which rely on knowing the signal parameters (symbol rate, modulation type, etc.)~\cite{prasad_lets_2011,hanna_maximizing_2016}.  To overcome the ignorance of these parameters, the dual path network uses   neural networks to predict the values  for the compensation. These networks are designed to enable joint estimation by maximizing the feature sharing among them.

	\subsection{Dual Path Design}
	The proposed network consists of two paths: a signal path consisting of linear operations that gradually restore the input signal, and a feature path made of  neural networks (NN) used to predict the parameters needed to recover the transmitted signal.   The  network is shown in Fig.~\ref{fig:dualPath}. The signal path is described as follows:
	It starts by compensating for the  frequency offset  using the network's   prediction $\widehat{f}_0$,   to obtain the vector $\widehat{\b{z}}_0$ defined as
	\begin{equation}
	\widehat{z}_0[k] = y[k] e^{-j 2\pi \widehat{f}_0 k}
	\end{equation}
	The noise of the signal is then attenuated using  the predicted  real filter taps $\b{w}_1$  and generates the vector  $\widehat{\b{z}}_1$ defined as
	\begin{equation}
	\label{eq:z1}
	\widehat{z}_1[k] = \widehat{z}_0[k] * w_1[k]
	\end{equation}
	where $*$ denotes the linear convolution operator.  At the end of the signal recovery, equalization is performed using the predicted complex filter taps $\b{w}_2$, to generate the vector  $\widehat{\b{z}}_2$ defined as
	\begin{equation}
	\label{eq:z2}
	\widehat{z}_2[k] = \widehat{z}_1[k] * w_2[k]
	\end{equation}
	Note that due to linearity, both filtering stages can be combined, and  a single set of predicted filters can perform both operations. However, we decided to keep the two filters separate similar to our prior work~\cite{hanna_combining_2020} to make training easier and for the network to extract features from two recovered signals instead of one.
	
	Feature Extractor NNs extract the features vectors $\b{g}_0$, $\b{g}_1$, $\b{g}_2$, and $\b{g}_3$ from the signals $\b{y}$,  $\widehat{\b{z}}_0$, $\widehat{\b{z}}_1$, and $\widehat{\b{z}}_2$ respectively. Each of these feature vectors is concatenated with the processed  previous features along the feature path,  which consists of 4 cascaded Feature Processing neural networks (NNs) as shown in Fig.~\ref{fig:dualPath}.  The predicted parameters used by the signal path $\widehat{f}_0$, $\b{w}_1$, and $\b{w}_2$ along with predictions of the timing offset  $\widehat{t}_0$, symbol duration $\widehat{\tau}$, and the modulation type  $\widehat{M}$ are obtained by their own dedicated estimation neural networks. The layer by layer description of the networks is discussed later in~ Section~\ref{subsec:net_arch}. The recovery of the symbols $\b{s}$ is performed using these parameters in post processing as described later in Section~\ref{subsec:dpn_post}.

	\subsection{Design Motivation and Considerations}
	In this section, we explain the motivation behind DPN design. The sequential stage by stage design enables later stages to benefit from the output of earlier stages. %
	The stage order is chosen based on the effect of the distortion on the signal time domain representation. A small frequency offset can significantly alter the signal in time domain  even at a high SNR, so it was considered first. Frequency recovery is also typically the first stage in classical demodulators~\cite{prasad_lets_2011}. For limited fading spread, the noise has a more pronounced effect on the signal, and it was considered as the second stage. However, other stage orders are possible; for instance, in our prior work~\cite{hanna_combining_2020}, the noise reduction stage was first due to a different choice of training data. The training data used in DPN is discussed later in Section~\ref{subsec:data}.
	
	Besides the 3 recovery stages, other DSP operations need to be performed to recover the symbols, however, there are design considerations that prevent integrating them into DPN. The first consideration is that all operations in DPN have to be differentiable to train it using gradient descent. The second consideration is that the output size needs to be determined only by the input size and should not depend on the network predictions. Although  there are known techniques to train NN that violate this consideration (commonly used in natural language processing), this consideration makes training easier.  The third consideration is that the network should not be penalized  for phase ambiguity. For example, if a signal $x(t)$ is a valid BPSK signal, $-x(t)$ is also a valid BPSK signal. Since the network does not have sufficient information to determine the correct one, it should not be penalized.

	These considerations make it hard to design a network that uses DSP modules to output the $N_s$ symbols $\b{s}$ for several reasons; (1) the relation between the signal length $N_r$ and $N_s$ depends on the number of samples per symbol $\tau$, which is unknown and varies from one training sample to the other, (2) the number of possible values of each output symbol is also a variable that depends on the modulation order, (3) considering phase ambiguity, there are several possible valid values of the sequence symbols. So instead of designing a network that outputs $\b{s}$ directly, we design a network that recovers the transmitted signal and estimates the parameters needed to decode $\b{s}$, which is done in post-processing. %

	\subsection{Design Merits and Drawbacks}
	The proposed DPN design combining  neural networks and signal processing has several merits:
	\begin{enumerate}
		\item \textit{Predicting using restored signals}: The less the distortions in the signal (higher SNR, smaller frequency offsets, etc.), the better the predictions that neural networks obtain~\cite{hauser_signal_2017}.  DPN design gradually restores the transmitted signal and uses the restored signal in the later predictions. This should lead to improved predictions.
		\item \textit{Incremental Feature Combining}: Each prediction made by the network has access to all features from the previous stages. For instance, the modulation classification NN, through the feature path can leverage all the previous features $\b{g}_1$, $\b{g}_2$, $\b{g}_3$, and $\b{g}_4$ not just $\b{g}_4$. The combining enables any prediction NN block to reuse  relevant features from the previous stages. It also works as a backup in case a signal reconstruction stage leads to degradation in the signal due to misprediction, which might occur at low SNR.
		\item \textit{Compatibility with Existing DSP Methods}: The DPN outputs are frequency estimates, filter taps, and timing estimates. All these parameters have a clear interpretation and can be reused using existing signal processing techniques. For instance, if the received signal is too long, DPN can be applied to obtain estimates on a short portion of the signal. The estimates like frequency offset can be directly applied to the remaining signal. In case there exists a time varying frequency drift, a frequency offset tracking algorithm can be applied to DPN's estimate.
	\end{enumerate}
	All three claimed merits are verified in the results section. To verify the first two merits, we consider two alternative designs of DPN, which are described later in Section~\ref{subsec:alternative_arch}.

	The design of DPN also has several drawbacks. Since the signal path consists only of linear operations, DPN can only perform linear reconstruction of the signal, that is the reconstructed output $\widehat{\b{z}}_2$ is obtained using linear operations on the input $\b{y}$. This is in contrast with using a fully neural network design which can approximate non linear relations between its input and output. However,   non linear distortions are typically mitigated in the radio hardware design, and communication systems are typically modeled using linear operations. The second drawback is that training DPN requires   different types of data for training. DPN requires the  true value of many signal parameters like frequency and timing offsets in addition to $\b{y}$. However, these parameters are easy to obtain in a simulator. To train DPN with an over the air capture, a long preamble can be used to precede known transmitted signals. Using existing signal processing techniques, the channel, the frequency,  and timing offsets can be estimated to obtain the required training data.
	
	These previous merits and drawbacks are specific to the proposed design. DPN also inherits some of the merits and drawbacks of deep learning. DPN  predicts using the weights calculated in training  instead on relying on manually designed features. This data driven approach makes it relatively easy to support another modulation type for instance at the cost of lacking an interpretation of what is being learned.

	\subsection{Training Data and Method}
	\label{subsec:data}
	\begin{table}
		\renewcommand{\arraystretch}{1.5}
		\caption{Loss Functions \label{tbl:loss}}
		\centering
		\begin{tabular}{|c|c|}
			\hline
			Equation \\	\hline
			$L_1 = |f_0 - \widehat{f}_0| $ \\ \hline
			$L_2 =\frac{1}{N_r} (\widehat{\b{z}}_1^H\widehat{\b{z}}_1  + \b{z}^H_1 \b{z}_1 - 2 |\widehat{\b{z}}_1^H  \b{z}_1| )$ \\ \hline
			$L_3 =\frac{1}{N_r} (\widehat{\b{z}}_2^H\widehat{\b{z}}_2  + \b{z}^H_2 \b{z}_2 - 2 |\widehat{\b{z}}_2^H  \b{z}_2| )$ \\ \hline
			$L_4 = (\widehat{t}_0-t_0)^2  $ \\ \hline
			$L_5 = (\widehat{\tau}-\tau)^2$ \\ \hline
			$L_6 = \text{crossentropy}(M,\widehat{M})$ \\ \hline
		\end{tabular} 
	\end{table}
	
	DPN is trained using  stochastic gradient descent. The labeled data used in training consists of received signals $\b{y}$ along with their corresponding values $f_0$, $\b{z}_1$,  $\b{z}_2$,  $t_0$,  $\tau$, and  $M$, where $\b{z}_1$ is the noise free received signal 
	\begin{equation}
	z_1[k]= \int_{-\infty}^{\infty} x(\gamma) h(t_k -\sigma) d\gamma 
	\end{equation}
	and  $\b{z}_2$ is the noise free and equalized signal
	\begin{equation}
	z_2[k]=  x(t_k) 
	\end{equation}
	Since, we are including the values of $f_0$, $\b{z}_1$,  $\b{z}_2$, the reconstruction stages are trained in a supervised manner.  To evaluate the feasibility omitting the signal ground truth in blind decoding  similar to~\mbox{\cite{hanna_asilomar_2017,oshea_radio_2016,shang_dive_2020,yashashwi_learnable_2018}}, we consider an unsupervised-reconstruction alternative architecture as discussed later in Section~{\ref{subsec:alternative_arch}}.
	Other choices of the training data are possible. For instance, instead of using  $\b{z}_1$ and  $\b{z}_2$, we could have designed a low pass filter $\b{w}_1$ and an equalization filter $\b{w}_2$ and used them for training. Since there are many possible ways to design these filters, instead of forcing the network to a specific design, we opted for only including the desired outputs  $\b{z}_1$ and  $\b{z}_2$ in the training. Also instead of using $f_0$  for training, we could have used a reference signal with no offset similar to what we did in our prior work~\cite{hanna_combining_2020}, but explicitly using $f_0$ was found to provide a better reconstruction. %
	Since we are providing the ground truth along the signal path,  and to avoid confusing the estimation networks,  gradients were prevented from propagating along the signal path according to the gradient stops shown in Fig.~\ref{fig:dualPath}.

	Since the outputs are heterogeneous, each output has its own loss function as defined in Table~\ref{tbl:loss}. For the modulation type, the classifier was trained using the categorical crossentropy loss. For the frequency and timing predictions, we  used the mean absolute error and the mean square error respectively. A natural choice for $\b{z}_1$ is the mean squared error (MSE) loss defined as $\frac{1}{N_r}\|\b{z}_1 - \widehat{\b{z}}_1\|^2 =  \frac{1}{N_r} (\widehat{\b{z}}_1^H\widehat{\b{z}}_1  + \b{z}^H_1 \b{z}_1 - 2 \widehat{\b{z}}_1^H  \b{z}_1 )$, where $(\cdot)^H$ denotes the hermitian operator. However, this choice penalizes constant phase offsets, and thus violates our design consideration for phase ambiguity. To avoid penalizing constant phase offsets, we used the modified loss shown in Table~\ref{tbl:loss} for $L_2$ and $L_3$. It is easy to verify that $\widehat{\b{z}}_1$ and $\widehat{\b{z}}_1 e^{j\phi}$ for any phase $\phi$ would yield the same loss. Note that the modified loss used in $L_2$ and $L_3$ is still sensitive to residual frequency offsets  $(\widehat{f}_0-f_0)$, which are time varying phase offsets.
	
	To support the feature combining along the feature path,  DPN is trained simultaneously as a single network with a loss $L=\sum_{k=1}^6 c_k L_{k}$, for some weights $c_k$. This implies that the gradients for a given NN backpropagate from all the subsequent outputs. For instance, the weights of the first feature extractor (generating $\b{g_0}$) are trained using  gradients from all the outputs  not just  $\widehat{f}_0$.  This makes the weights of each NN along the feature path optimized to minimize the total loss and not the loss of  a specific output.

	\subsection{Scalable Neural Network  Architecture }
	\label{subsec:net_arch}
	\begin{figure}[t!]
		\centerline{\includegraphics[scale = 1]{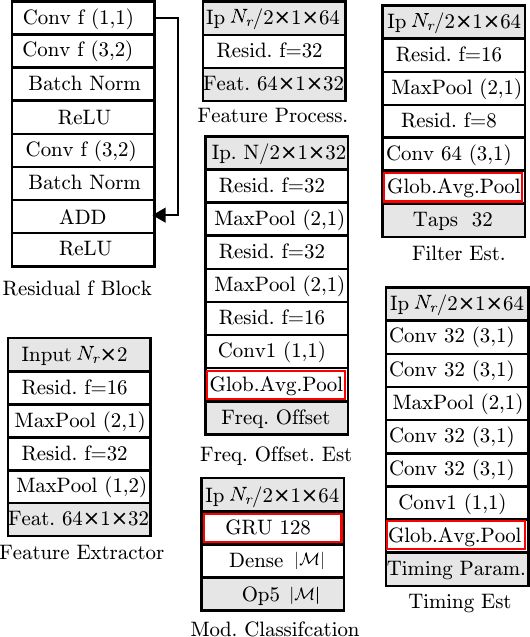}}
		\caption{The layer by layer description of the NNs in DPN.}
		\label{fig:dualPathLayers}
	\end{figure}
	We start by describing the building blocks of the network. The entire network consists of a combination of convolutional networks and recurrent neural networks. This choice was not only based on their outstanding results in modulation classification~\cite{hong_automatic_2017,oshea_over--air_2018}, but also because their number of trainable parameters is independent from the input length~\cite[Ch.10]{goodfellow_deep_2016}. A convolutional layer is based on the convolution operation, in which the number of trainable filter weights is independent of the input size. Similarly, a recurrent neural network uses the same weights for any number of time steps~\cite[Ch.10]{goodfellow_deep_2016}.  %

	Then, we describe each network in details. The convolutional layers used in this work were arranged as residual blocks. Residual blocks improve the gradient flow during the training of deep networks~\cite{he_deep_2015} and were proposed for modulation classification~\cite{oshea_over--air_2018}.  
	The neural network blocks layer by layer descriptions are shown in Fig.~\ref{fig:dualPathLayers}.
	Both timing offset and symbol duration prediction used the same architecture referred to as Timing Est. The noise reduction filter used the Filter Est shown in the same figure. The fading filter used has complex taps, hence needs twice the number of real outputs. It was obtained by modifying  the Filter Est block to have  two 32 sample outputs by duplicating the layers following the last residual block. The modulation classification block consists of a gated recurrent unit (GRU) network, which is a type of recurrent network followed by a dense layer.
	
	To support variable length signals, all layers generating the network outputs either use global average pooling or recurrent neural networks. Global average pooling  calculates the average along the time dimension.  Hence, it does not have any trainable weights and is independent of its input time dimension.  This is in contrast with most existing works, which rely on a dense network after the convolutions with the number of trainable weights dependent on the time dimension. The output of the recurrent network is obtained from the last time step and hence its output dimension is independent from the time dimension. The  layers  generating outputs independent of the time dimension are highlighted in red in Fig.~\ref{fig:dualPathLayers}.
	
	DPN has a total of 233,419 trainable parameters. This number of parameters is independent of the input length $N_r$. No matter how long the input is, the number of weights is constant. However, the fact that the same weights can be reused for the different lengths does not imply that the performance will generalize across different lengths. We will explore the effect of changing the signal input length on the performance in our evaluation.
	\subsection{DPN Implementation and Training Parameters}
	The code for DPN is publicly available\footnote{https://github.com/uclacores/dual\_path\_network}. DPN was implemented using the KERAS API for TensorFlow.  It was trained for 100 epochs, with early stopping occurring if the validation loss did not improve for 10 consecutive epochs. The best weights according to the validation loss are retained. Note that early stopping is used to determine when the network weights have converged and not to avoid overfitting. Overfitting is avoided by using realtime signal generation as discussed later in Section~\ref{sec:data_gen}.
	
	The weights for the losses $c_1$ to $c_6$ were set to 500, 1, 5, 2.4e-4, 4.8e-4,  1.0  respectively. These values were chosen to account for the relative importance of the parameters and the different magnitudes of the losses which depend  on the training labels values that are described later.  For instance   frequency offset loss ($L_1$) takes   values in the range of 1e-2, and has a large impact on blind decoding and modulation classification and hence was assigned a large weight $c_1=500$. In contrast, the  timing offset loss ($L_4$) takes values in the range of 1K (when using integer sample offsets) and hence was assigned a small  weight $c_4=2.4e-4$. The optimizer used for training is the ADAM optimizer with a learning rate of 0.001  and the gradients were clipped at a norm of 1.0 to avoid exploding gradients~{\cite{pascanu_difficulty_2013}}.
	
	\subsection{ Alternative Architectures}
	\label{subsec:alternative_arch}
	\begin{figure}[t!]
		\begin{center}
			\subfloat[Single Path\label{fig:single_path}]{\includegraphics[width=3.5in]{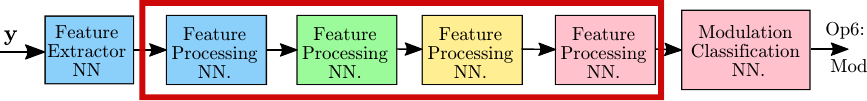}}  
			
			\subfloat[Separate DPN (SepDPN) \label{fig:sep}]{\includegraphics[width=3.5in]{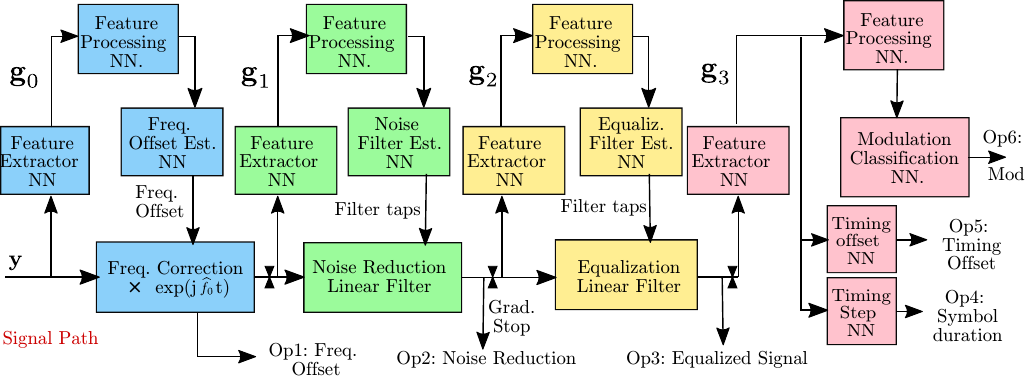}}  
		\end{center}
		\caption{Alternative DPN designs.}
		\label{fig:dpn_alt}
	\end{figure}
	
	We consider two alternative designs for DPN to verify some of the claimed merits and a third one to evaluate the benefit of supervised reconstruction. The first design aims to quantify the impact of restoring the signal on modulation classification. This design  omits the signal path entirely from DPN as shown in Fig.~\ref{fig:single_path} and is referred to as Single Path. If restoring the signal improves predictions as claimed, DPN should outperform Single Path in modulation classification.
	
	The second alternative design aims to verify the benefits of feature combining. To prevent feature combining, the feature path is disconnected as shown in Fig.~\ref{fig:sep} preventing feature combining.  Since the signal path contains gradient stops,  the gradients of one stage do not propagate to the previous stage and the stages are separate. This network referred to as Separate DPN (SepDPN) and is equivalent to training one stage, calculating the predictions, and using them to train the following stage.  Hence, in this design, the features $\b{g}_0$ are used exclusively for the prediction of $\widehat{f}_0$, and similarly for the later stages. If feature combining leads to an improvement in prediction as claimed, DPN should  outperform SepDPN. 
	
	The third alternative considers the effect of training of the signal reconstruction stages without the ground truth similar to what was proposed in~\mbox{\cite{hanna_asilomar_2017,oshea_radio_2016,shang_dive_2020,yashashwi_learnable_2018}} for modulation classification. The design is similar to DPN design shown in Fig.~\ref{fig:dualPath} after omitting the ground truth values of $f_0$, $\b{z}_1$,  $\b{z}_2$,  $t_0$, and $\tau$, leaving only the modulation type $M$.  To train the network, the gradient stops along the signal path were removed.  The timing networks were also  omitted because they cannot be trained without the ground truth. This alternative design, referred to as unsupervised DPN (unsupDPN), investigates whether unsupervised reconstruction is valid for blind decoding or not. It also investigates the impact of supervised reconstruction on modulation classification.

	\section{Signal Recovery and Symbol Decoding Algorithm}
	In this section, we discuss the benchmark algorithms for signal recovery used for comparison. We also discuss DPN postprocessing and  symbol decoding.
	\subsection{Benchmark Signal Recovery Algorithms}
	To evaluate the performance of DPN in signal recovery, we   contrast it with exiting DSP techniques. To that end, we consider two algorithms; the first one is a fully blind  recovery algorithm based on signal processing, the second is a genie algorithm that assumes all the unknown signal parameters are provided by a genie. It is important to note that there exist many signal processing algorithms for recovery whether blind or genie. These algorithms have a tradeoff between complexity and performance. The chosen algorithms were selected to be non-iterative and to rely on well known DSP techniques.

	\subsubsection{Blind DSP Reference Algorithm (DSP Ref)}
	\label{subsec:ref_dsp}
	The reference blind algorithm  used is based on the symbol recovery proposed in~\cite{rebeiz_energy-efficient_2014}.  It starts with  band segmentation stage aiming to obtain an initial estimate of the signal center frequency and bandwidth. It is performed using the Welch power spectral density as described in Appendix~\ref{ap:band_seg}.
	The initial estimates are refined by detecting the signal's cyclostationary features   as proposed in~\cite[Sec. III-B]{rebeiz_energy-efficient_2014}.   %
	The exact refinement procedures  are described in Appendices~\ref{ap:freq} and~\ref{ap:symb_rate} for center frequency and symbol rate  respectively.  
	After the estimation of $\tau$ and $f_0$, the timing offset $t_0$ is estimated using the Gardner symbol timing recovery algorithm~\cite{gardner_bpsk/qpsk_1986} as described in Appendix~\ref{ap:timing_off}. 
	The constant modulus algorithm (CMA) is then applied for blind signal equalization~\cite{qinghua_shi_blind_2012}  as described in Appendix~\ref{ap:blind_eq}.
	
	\subsubsection{Genie Algorithm}
	The genie algorithm assumes all the parameters which were used to generate the signal are known. It starts by correcting the frequency offset, then applying a low pass filter over its bandwidth, which are both assumed to be known. It uses the MMSE equalizer assuming that the channel is known as described in Appendix~\ref{ap:genie_eq}. The symbol recovery is performed using the true $\tau$ and $t_0$ using the same procedures as DPN output, which are described in Section~\ref{subsec:dpn_post}.
	
	\subsection{DPN Post Processing for Signal Recovery}
	\label{subsec:dpn_post}
	The signal output of DPN is given by $\widehat{\b{z}} = \widehat{\b{z}}_2$. This signal is already frequency compensated, noise reduced, and equalized by the network. The remaining processing that needs to be applied to this signal is the symbol recovery. This is done using the estimated timing offset $\widehat{t}_0$ and  symbol duration $\widehat{\tau}$. To apply the timing recovery, the signal is interpolated by a factor of $P$, larger than any value of $\tau$ of interest,  to get the signal $\widehat{\b{z}}_I$. The vector  $\widehat{\b{z}}_I$ is padded at the start with   $\widehat{t}_0 P$ zeros to correct for the time offset and then sampled every $P \widehat{\tau}$.
	
	\subsection{Symbol Decoding}
	\label{susbec:symbol_decoding}
	The symbol decoding procedure is applied to the recovered symbols  to evaluate the symbol error rate (SER) for the linear modulations. This same procedure is applied to the symbols calculated from  all of the previous approaches (DPN, blind DSP, and genie). The symbols are decoded  symbol per symbol, using a minimum distance euclidean receiver.  A decision aided phase recovery loop uses the predicted symbol to correct the phase of the next symbol. This loop helps reduce the impact of any minor residual frequency errors. For evaluation purposes,  the first symbol is assumed to be known and is used to estimate the phase to address phase ambiguity.  The step-by-step procedure is described in Appendix~\ref{ap:symb_decod}. 
	
	Although this approach is simple, its main disadvantage is that it propagates errors. A decision error in one symbol  propagates to the remaining symbols. Since only the first symbol is used to estimate the phase in the evaluation, the initialization of the loop is subject to errors, which can propagate.  More sophisticated receivers can be developed, however, we chose a simple receiver that would work for all linear modulation types.  Since this receiver is common to all approaches, it should not create a bias in the evaluation.

	\section{Data Generation and USRP Capture}
	\label{sec:data_gen}
	\begin{figure}[t]
		\centerline{\includegraphics[scale=1]{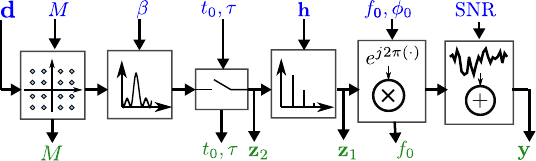}}
		\caption{Flow graph for generating samples showing on top the input parameters and the bottom the outputs used for training.}
		\label{fig:generator}
	\end{figure}
	
	To train and test DPN, we generated a dataset consisting of signals with different data, modulation types, symbol rates, timing and frequency offsets, phase, channel impulses, and SNRs. While public datasets exist for modulation classification, they do not contain all the required signal labels, and hence are not suitable for blind decoding.    In the dataset, each signal is generated according  to the flow graph shown in Fig.~\ref{fig:generator}. Random data $\b{d}$ is generated and modulated using modulation type $M$ selected from the set of single carrier modulations  $\mathcal{M}$. If $M$ is a linear modulation, the output is upsampled by an integer factor $N_{up}$ and pulse shaped with a root-raised-cosine filter with a roll-off factor $\beta$. To simulate timing offsets, the first $t_0 N_{up}$ (rounded to the nearest integer) samples are removed  and the signal is downsampled  by a factor of  $\left\lfloor N_{up}/\tau \right\rfloor$. While removing  $t_0 N_{up}$ might make the first symbol unrecoverable, it is more realistic for a blind system as it emulates capturing an ongoing transmission. Since the number of removed samples and the downsampling are integers,  the number of possible values of $t_0$ and $\tau$ is discrete and depends on $N_{up}$ and the range of values of $\tau$. Multipath fading is simulated using convolution with random complex fading taps  having a delay spread~$\sigma$. Then frequency and phase offsets, $f_0$ and $\phi_0$, are applied, and Gaussian noise is added to model different SNRs.
	
	The dataset consists of signals generated with  parameters  uniformly sampled from the ranges shown in Table~\ref{tbl:datasets}. The upsampling factor was chosen to be equal to $N_{up}=64$, making the number of  possible values  of $t_0$ and $\tau$ equal to 64 and 14, respectively.
	The channel  $\b{h}$ has  3   non random zero complex taps  having  a  Rayleigh magnitude and uniform phase. The 3 taps are spaced uniformly within the fading spread.  Since  single carrier modulations are more commonly used in narrowband systems with limited fading, the fading spread is assumed to be  limited within one symbol duration.   Unless otherwise stated, the signal length  is given by $N_r=1024$. Note that in Table~\ref{tbl:datasets}, the assumption of a sampling duration of 1 second  is only made to normalize  parameters and simplify the notations as stated in the system model. For a 20MHz  sampling rate (50ns duration), for instance,  all durations in the table should be multiplied by 50ns and the frequency offset by 20MHz.  
	
	Typically, a fixed dataset is used in training,  and data augmentation is performed to avoid overfitting. Since our dataset is generated using simulations, instead of fixing the size of the training data and using augmentations, we generated the signals in real-time during training.  Real-time signal generation, while eliminating overfitting,  requires optimizing the generator execution time to avoid slowing down the training significantly. For training DPN, in each epoch 800K new signals are generated, the total training data for 100 epochs is 80M signals. However, note that DPN can work with a smaller fixed dataset with appropriate regularization and data augmentation similar to any other network. Also, all the networks used for comparison in this work used  real-time sample generation. The validation and test sets  contain a fixed 100K samples.  Both sets use the same parameters given in Table~\ref{tbl:datasets}, except that the test set SNR was discretized from 0 to 20dB at  5dB steps for a more convenient result visualization.
	
	\begin{figure}[t]
		\centerline{\includegraphics[scale=.85]{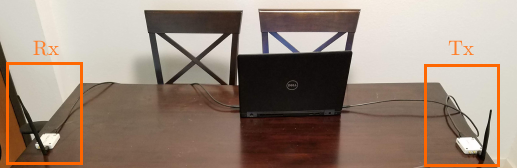}}
		\caption{Over-the-air capture setup using USRPs B205 mini.}
		\label{fig:usrp}
	\end{figure}
	
	To validate DPN's performance beyond simulations, we collected an over-the-air capture using software-defined-radios (SDRs). The capture was performed indoor using two USRP B205 mini SDRs~\cite{ettus_research_usrp_nodate} in a  line-of-sight channel as shown in Fig.~\ref{fig:usrp} using a center frequency of 915MHz  a sampling rate of 1M samples per second ($\tau_0 = 1\mu s$). The receiver center  frequency was intentionally offset by 5KHz (0.005Hz at $\tau_0=1$) to emulate the lack of prior agreement, in addition to the offsets due to the USRP oscillator accuracy having a rang e of   $\pm 2$ppm~\cite{ettus_research_usrp_nodate} ($\pm 1.83$KHz  at 915MHz).  The Tx USRP sent 100K signals  following the same modulation types and normalized symbol durations (samples per symbol)  in Table~\ref{tbl:datasets} over a period of 3 minutes. The Rx USRP captured  the signals, which were  isolated and matched to the corresponding transmitted signals in post processing. This process was repeated three times with different signal amplitudes to obtain 3 captures with estimated SNRs equal to  8,14, and 20dB. The SNR was estimated by measuring the received signal power and dividing it with measured power with no active transmissions. Hence, the SNR does not account for quantization noise, which is  more significant for smaller signal amplitudes. The received signals, transmitted symbols, and the modulation types form the  over-the-air test dataset at a given SNR. %

	\begin{table}[t!]

		\renewcommand{\arraystretch}{1.5}
		\caption{Dataset Description \label{tbl:datasets}}
		\centering
		\ifdefined \singleCol
		\begin{tabular}{|p{1.2 in}|c|p{4.5 in}|}
			\else
			\begin{tabular}{|p{0.8 in}|c|p{2 in}|}
				\fi
				\hline
				Desc. & Par. & Range \\	\hline
				Modulation Types  &$M$& \{OOK, ASK4, ASK8,  
				OOK,  ASK4, ASK8, BPSK, QPSK, PSK8, PSK16, PSK32, APSK16, APSK32, APSK64, QAM16, QAM32, QAM64, GMSK, CPFSK\} \\ \hline
				Sampling Period & $\tau_0$& 1 \\ \hline
				Frequency Offset  & $f_0$& 	$[-0.01,0.01]$ \\ \hline
				SNR (dB) & SNR & 	$[0,20]$ \\ \hline
				Timing Offset  & $t_0$ &  $[0,1]$ \\ \hline
				Symbol Duration & $\tau$& 	$[4,16]$ \\ \hline
				Pulseshape rolloff & $\beta$& 	$\{0.15,0.25,0.35\}$ \\ \hline
				Phase Offset & $\phi_0$& 	$[0,2\pi]$ \\ \hline
				Delay Spread  & $\sigma$ &  [0,$\tau$] \\ \hline
			\end{tabular} 
		\end{table}

		\section{Results}
		
		\subsection{Signal Recovery}
		\begin{figure*}[t!]
			\begin{center}
				\subfloat[Carrier Frequency Offset\label{fig:eval_freq}]{\includegraphics{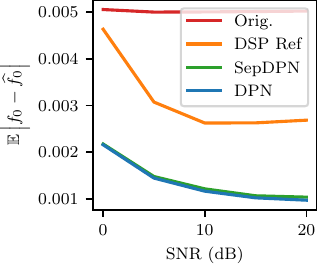}}  \hspace{0.5mm}
				\subfloat[Symbol duration \label{fig:eval_sps}]{\includegraphics{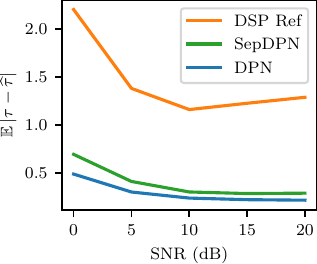}}  \hspace{0.5mm}
				\subfloat[Timing Offset \label{fig:eval_timing_off}]{\includegraphics{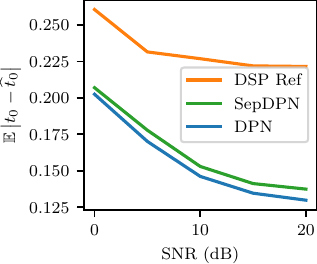}}  
			\end{center}
			\caption{The mean absolute error  of DPN in parameter estimation.}
			\label{fig:params}
		\end{figure*}
		
		\begin{figure}[t]
			\centerline{\includegraphics[scale=1]{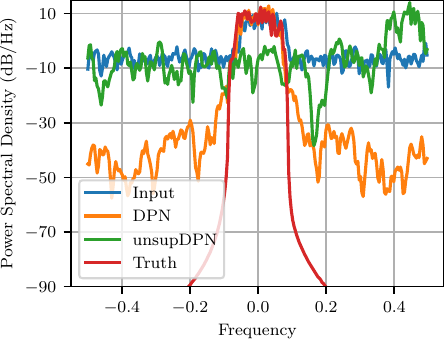}}
			\caption{The frequency representation of an example input signal, the signal recovered by DPN, and the ground truth.}
			\label{fig:psd}
		\end{figure}
		
		We start by evaluating the   performance of DPN, SepDPN, and DSP Ref  in recovering the transmitted signal.  While signal recovery is not our main objective in this paper, it helps verify the merits of DPN and explain the blind decoding and modulation classification results. We start by evaluating the mean absolute error for the frequency offset $\widehat{f}_0$,  timing offset $\widehat{t}_0$, and the symbol duration $\widehat{\tau}$ as  a function of the SNR.   The mean absolute error  for  $\widehat{f}_0$ is calculated using $\mathbb{E}|f_0-\widehat{f}_0|$  on the test dataset where $\mathbb{E}$ denotes the expectation.
		
		The results are shown in Fig.~\ref{fig:params}. In Fig.~\ref{fig:eval_freq}, we show the original frequency offset (Orig.), which is the mean absolute value of a uniform variable from [-0.01,0.01] as stated in Table~\ref{tbl:datasets}. We can see that for all methods the estimation error decreases as the SNR gets larger and that the residual frequency offset is lower than the original offset. Comparing the two variations of DPN, both give almost identical frequency estimation errors. This is expected since both approaches for the first frequency offset stage use only the features from the input signal. The curve for unsupDPN was omitted as it provided values above 55, which are obviously erroneous. Since unsupDPN is trained to reduce the modulation classification loss, the predicted frequency is an arbitrary value that reduces this loss with no signal processing meaning.
		
		The DSP Ref algorithm using the  parameters  in the Appendices gives a higher estimation error than DPN. It is important, however, to note that DPN has the advantage of having prior information about the signals from the training data,  unlike  DSP Ref. The purpose of comparing to the blind DSP algorithm is just to provide a baseline reference for comparison. Each approach is considered in the way it is typically implemented and we do not claim to provide an absolute comparison between DSP algorithms in general and deep learning.  This trend of results between DSP Ref and the DPN variations continues for the symbol duration and timing offset estimation. It is  noteworthy that DSP Ref uses the symbol duration in the calculation of the timing offset. A mistake in the former will  lead to a mistake in the latter, which would lead  to a high  symbol error rate in the same signal. 
		
		Then, we consider the results for the later stages of DPN; symbol duration in  Fig.~\ref{fig:eval_sps} and timing offset in Fig.~\ref{fig:eval_timing_off}.  We see that the proposed DPN with feature sharing performs better than the separately trained DPN, in contrast to the first stage's where the results were close. This is attributed to the fact the symbol duration and timing estimation networks of the proposed DPN, due to being at the end of the network,  benefit from all the  extracted features. SepDPN, on the other hand, only uses the features from the last stage. This shows that feature sharing can lead to an improvement in estimation. 
		
		The fact that the predicted  estimations are close to the true values makes them reusable by signal processing algorithms. For instance, the frequency offset obtained can be used as an initial value for any carrier frequency offset tracking algorithm~\cite{prasad_lets_2011}. The symbol duration and offset can be used to initialize a symbol timing recovery algorithm~\cite{prasad_lets_2011}. This can be useful in scenarios where a long signal is received at a high sampling rate and processing the entire signal with neural networks is not computationally feasible.
		
		Lastly, we  show  the recovery results on the example signal visualized in Fig.~\ref{fig:dualPath}. For the BPSK signal having $\tau = 8$ and an SNR = 5dB and fading spread $\sigma=0.3$, in Fig.~\ref{fig:psd}, we show  the power spectral density (PSD) of the input $\b{y}$, the recovered signal $\widehat{\b{z}}$ from DPN, unsupDPN, and the true clean signal $\b{z}$.  These signals in time domain are visually hard to compare because of phase offsets, that is why we used the PSD. From that Figure, we can see the filter taps  predicted by DPN attenuated the noise by over 25dB  and made the output more similar to the truth. This shows that using only training data, DPN can predict filters that partially recover the transmitted signals.  In contrast, the recovered signal from unsupDPN is arbitrary and hence has no value in blind decoding. This shows the necessity of including the ground truth for blind decoding.   Instead of evaluating the signals $\widehat{\b{z}}_1$ and $\widehat{\b{z}}_2$ over the entire dataset, which is not straight forward due to residual frequency errors, we move on to the  symbol decoding evaluation. Symbol  decoding implicitly evaluates the recovery stages and is our objective in this paper.

		\subsection{Blind Symbol Decoding}
		\begin{figure}[t]
			\centerline{\includegraphics[scale=1]{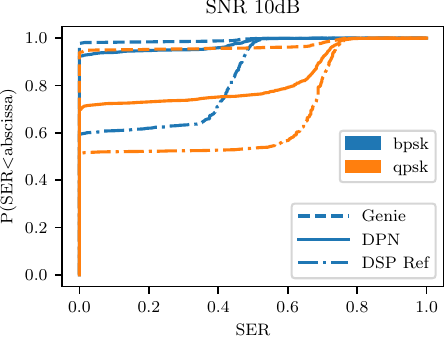}}
			\caption{The CDF of the symbol error rate (SER) calculated for each signal at SNR=10dB of different approaches.}
			\label{fig:ser}
		\end{figure}
		\begin{figure}[t]
			\centerline{\includegraphics[scale=1]{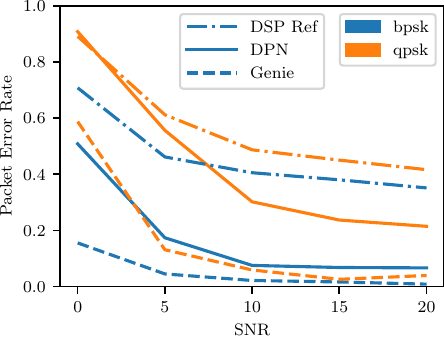}}
			\caption{The packet error rate of different approaches plotted against SNR.}
			\label{fig:per_snr}
		\end{figure}
		Then, we evaluate DPN performance in blind symbol decoding. Blindly decoding symbols is challenging since the signal parameters are not known by the blind receiver and no known preambles are assumed.   Even at a high SNR, synchronization errors can lead to a high symbol error rate. Additionally, the symbol decoding procedure used in this work is relatively simple and not optimized for any specific modulation type. All these factors make obtaining a low symbol error rate (SER) as expected in a typical communication system impractical. A completely blind receiver also would not expect to obtain the same performance as a receiver with protocol knowledge. 
		
		For these reasons, we limit  our results in this section to BPSK and QPSK modulations.  We calculate the symbol error rate (SER) for each signal and plot the CDF of SER over all the test signals when using DPN, Genie, and DSP Ref for signal recovery in Fig.~\ref{fig:ser} at SNR = 10dB. DPN obtains   lower SER  for a larger fraction of the signals compared to Ref DSP. This follows from the fact that DPN had  better  results in signal recovery.   As expected Genie performs better than DPN since it has access all the signal and channel parameters. Since BPSK is a lower order modulation, it can tolerate a larger  synchronization errors and the gap between Genie and DPN is smaller for BPSK than QPSK. Notice that the CDF has a jump near SER 0.5 for BPSK and SER 0.75 for QPSK. These values are equivalent to a random guess (0.75 and not 0.5 for QPSK since we are using SER and not bit error rate). This jump is due to either a significant mistake in parameter estimation or an error in one symbol propagating to the remaining symbols in a given signal. From these results, we see that each signal is either decoded almost entirely correctly or not. 
		
		Based on this observation, we consider each signal to be a packet and consider the packet error rate (PER), equivalent to P(SER=0), as a function of SNR for the same modulation types in Fig.~\ref{fig:per_snr}.  The results follow the same trend with Genie having the lowest SER followed by DPN and then DSP Ref. Since this curve is for packet error and  there are between 64 and 256 symbols per packet, a random guess would yield a PER close to one.
		Despite that the problem of blind symbol decoding is challenging, DPN performance, while not great in absolute terms, is better than the blind DSP Ref and the  PER gap between it and Genie gets as low as 5\% for BPSK.

		\subsection{Modulation Classification}
		\begin{figure}[t]
			\centerline{\includegraphics[scale=1]{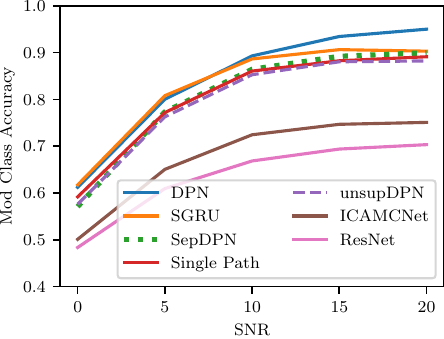}}
			\caption{The modulation classification of DPN, its alternatives SepDPN, and Single Path, and networks from the literature.}
			\label{fig:mod}
		\end{figure}
		\begin{figure}[t]
			\centerline{\includegraphics[scale=0.8]{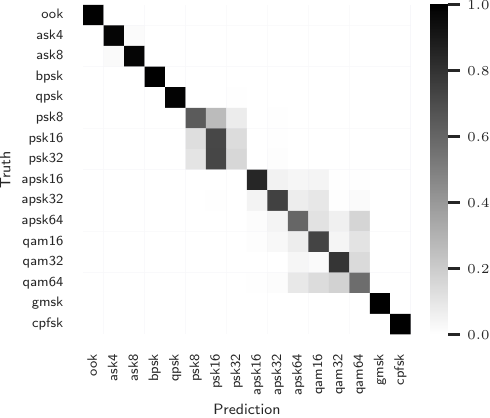}}
			\caption{The confusion matrix of DPN at SNR of 5dB.}
			\label{fig:conf}
		\end{figure}
		
		In this section, we consider the modulation classification performance of DPN and its variations (SepDPN, UnsupDPN, and Single Path). We also consider networks from the literature, namely the Stacked GRU (SGRU)~\cite{hong_automatic_2017} and the ResNet~\cite{oshea_over--air_2018}, and ICAMCNet~\cite{icamcnet_2020}. SGRU consists of recurrent networks and  ResNet and ICAMCNet are fully convolutional networks. All networks considered are trained with real-time sample generation using the same procedures as DPN. Note that  SGRU, ResNet, Single Path and unsupDPN   use only the modulation type as a label in contrast to  DPN and SepDPN  that additionally use the frequency offset, noise reduced signal, equalized signals, and timing information. For fairness, the modulation-only networks were allowed  up to 4 times the training epochs and up to 4 times the training data as DPN (up to 400 epochs and  320M training signals with early stopping).

		The results are shown in Fig~\ref{fig:mod}. The two networks giving the highest accuracy are DPN and SGRU. At SNR less than 10dB, both approaches yield almost the same accuracy. For SNRs larger than 10 dB,  SGRU performance starts to saturate and DPN performance improves surpassing SGRU by up to 5\%. This is explained by considering the signal recovery performance. At higher SNR, the signal recovery stages perform better, providing  signals with less distortions for the modulation classification. Signals with less distortions result in  better predictions. At lower SNR, the improvements to the signal along the signal path are more limited and do not lead to significant gains  compared to SGRU. Even though DPN and SGRU attain the same accuracy below 10dB, DPN can  be used for blind decoding,  and is about twice faster in inference compared to SGRU as discussed later.
		The fully convolutional networks ICAMCNet and ResNet gave a lower accuracy than SGRU and DPN that use recurrent networks for modulation classification for this dataset having variable samples per symbol. In Fig.~{\ref{fig:conf}}, we show the confusion matrix of DPN at 5dB SNR, from which we see that the  confusion occurs mostly in high order modulations of the same type. Note that  many improved deep learning layers and architectures were proposed for modulation classification in the literature. Our focus in this work is on combining  DSP with deep learning and not just using improved deep learning modules for higher classification accuracy. Improved deep learning modules can easily be integrated within the dual path architecture. Hence, we focus more on interpreting  DPN's design performance by comparing it with its alternative architectures.  %
		
		\begin{table}[t!]
			\renewcommand{\arraystretch}{1.5}
			\caption{Impact of each stage on Mod. Class. Accuracy } \label{tbl:ablation}
			\centering
			\begin{tabular}{|c|c|c|c|c|c|c|}
				\hline
				Zeroed Loss & None &$L_1$ & $L_2$  & $L_3$& $L_4$  & 	$L_5$      \\	\hline
				Mod. Class.  & 83.7\%& 80.5\% & 81.6\%  & 83.4\% & 83.11\% & 82.8\% \\	\hline
			\end{tabular} 
		\end{table}
		
		To better understand the factors leading to DPN's results, we consider its variations. First, we see that DPN outperforms Single Path. Since Single Path consists of the exact deep-learning-based feature path of DPN, the improved results of DPN are not solely attributed to the design of the feature path. The DSP signal recovery leads to a significant  accuracy improvement.To further understand the impact of each recovery stage individually on the classification, several instances of  DPN were trained while setting the weight of one of the losses $L_1$ to $L_5$ to zero. The obtained accuracies are shown in Table~III. From this table, we see that eliminating the frequency correction by zeroing $L_1$ has the largest impact on classification, since the dataset uses high order PSK signal which are sensitive to phase shifts due to frequency errors. On the other hand, the timing offset loss $L_4$ has the smallest impact because of its limited impact on the signal  when using more than 4 samples per symbol.  Comparing DPN to SepDPN, which lacks features sharing, DPN is better at identifying the signal modulation type. Since the only advantage DPN has over SepDPN is the connection of the feature path,  feature sharing that occurs along this path also contributes to the improved performance. Hence, by leveraging signal reconstruction and feature sharing, DPN design  improves modulation classification by up to 5\%. For unsupDPN, we see that the accuracy is also about 5\% less than DPN. This result shows that including the reconstruction ground truth  has the added benefit of improving the modulation classification. The similarity between the results of unsupDPN and Single Path indicates that  unsupDPN was not able to leverage the reconstruction modules to improve the accuracy.
		
		\begin{table}[t!]
			\renewcommand{\arraystretch}{1.5}
			\caption{Training and Inference Times \label{tbl:train_infer_time}}
			\centering
			\begin{tabular}{|c|c|c|}
				
				\hline
				Net. & Training (s/epoch)  & Inference ($\mu s$/signal)  \\	\hline
				DPN & 578.82 &  736 \\	\hline
				SGRU  & 339.22 &  1530 \\	\hline
				ResNet & 214.52 &  168 \\	\hline
				ICAMCNet & 227.60 &  135 \\	\hline
			\end{tabular} 
		\end{table}

		Lastly, the training and inference times of the neural networks are shown in Table~IV. The training time is calculated as the average of training time per epoch. The inference time is calculated by averaging the time to make predictions per signal over the test set. The server used for calculations has an Intel Core i9-9920X (12 Cores, 3.50 GHz), an RTX 2080Ti GPU, and 128GB of RAM. From Table~IV, DPN is slower to train than the other networks, since it has 6 outputs and thus requires more slow memory operations to transfer the data to the GPU for training. Note that since we are using real-time signal generation, the training time includes the overhead signal generation. In terms of inference time per signal (1024 IQ samples), DPN is slower than the fully convolutional networks (ResNet, ICAMCNet), however it is about twice faster than the SGRU. Recurrent neural networks (like SGRU and the modulation classification NN of DPN) have data dependencies among their different units, which makes them slower compared to convolutional networks. DPN consists mostly of convolutional layers except for one  GRU layer much smaller that those  in SGRU, and hence is faster.
		
		\subsection{Over-the-Air USRP Evaluation}
		\ifdefined \singleCol
		\begin{figure*}[t!]
			\begin{center}
				\subfloat[Modulation Classification\label{fig:usrp_mod}]{\includegraphics{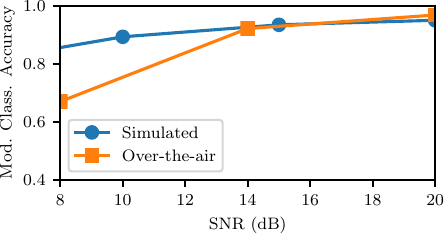}}  \hspace{1mm}
				\subfloat[PER \label{fig:usrp_per}]{\includegraphics{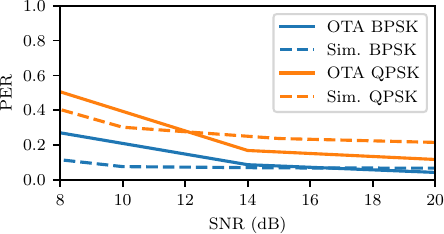}}  \hspace{1mm}
			\end{center}
			\caption{Comparison between simulation and over-the-air results.}
			\label{fig:usrp_res}
		\end{figure*}
		\else
		
		\begin{figure}[t!]
			\begin{center}
				\subfloat[Modulation Classification\label{fig:usrp_mod}]{\includegraphics{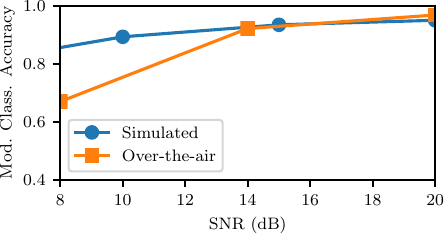}}   \\
				\subfloat[PER \label{fig:usrp_per}]{\includegraphics{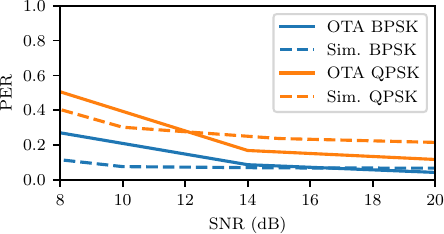}}  \hspace{1mm}
			\end{center}
			\caption{Comparison between simulation and over-the-air results}
			\label{fig:usrp_res}
		\end{figure}
		\fi
		To validate our signal model, we evaluate DPN, trained with the simulated data,  using the over-the-air test dataset without any retraining.
		The obtained results are shown for modulation classification in Fig.~\ref{fig:usrp_mod} and packet error rate in Fig.~\ref{fig:usrp_per}. From these Figures, we see that the OTA results are close to the simulated datasets for the  14dB and 20dB captures\footnote{The PER of the simulated QPSK signal is higher above 14dB because the simulated signals can have larger frequency offsets (up to 0.01Hz) compared to 0.005Hz for the OTA capture.}.  This result verifies our signal model and  DPN's performance at this range of SNR.  At  8dB SNR, DPN results degrade with lower modulation accuracy and higher PER.  This degradation is attributed to quantization errors, which become more significant in weaker signals and has not been accounted for in the SNR estimation, along with other hardware imperfections (amplifier non-linearity, IQ imbalance, etc). These imperfections, which are typically not modeled, are known to impact RF deep learning classifiers~\cite{clark_iv_training_2020}.  While retraining using the OTA capture can improve performance on this specific radio capture~\cite{oshea_over--air_2018}, it does not guarantee that the results will generalize to different radio hardware or deployment environment and thus we do not consider it. A more in depth study of the performance trade-offs of using simulated and over-the-air  data for training was performed in~\cite{clark_iv_training_2020}.  
		
		\subsection{Variable Length Evaluation}
		\begin{figure*}[t!]
			\begin{center}
				\subfloat[Frequency Offset\label{fig:freq_pkt}]{\includegraphics{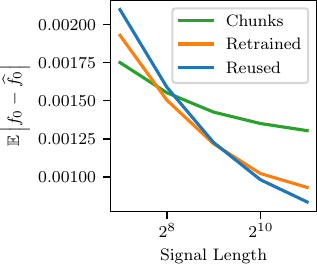}}  \hspace{0.5mm}
				\subfloat[Symbol duration\label{fig:sps_pkt}]{\includegraphics{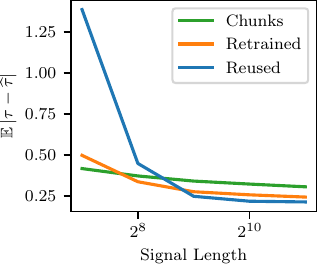}}  \hspace{0.5mm}
				\subfloat[Modulation Classification \label{fig:mod_pkt}]{\includegraphics{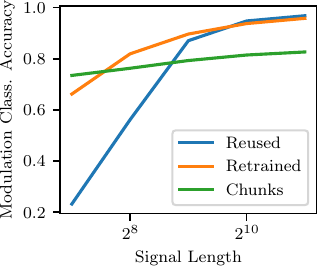}}  %
			\end{center}
			\caption{Comparison of different approaches to scale DPN; dividing the signal into chunks, reusing the weights trained for $N_r=1024$, retraining using different input lengths.}
			\label{fig:params_pkt}
		\end{figure*}
		As discussed earlier, DPN architecture supports using variable input lengths using the same weights. Our previous results were only for signals of length  $N_r = 1024$. In this section,  we evaluate the ability of DPN to generalize to lengths different than the training signals. We consider the lengths  $N_r \in \{128,256,512,1024,2048\}$  for testing with 10000 signals generated at each length using a 20dB SNR.  Three approaches are considered; in the first approach, we reuse the same weights  $\theta_1$ trained on signals having $N_r=1024$ without any modifications.  In the second approach, we retrain the network starting from $\theta_1$ on signals of different lengths to obtain the weights $\theta_2$. The retraining was done for 20 epochs such that the signals in each training batch had a length chosen randomly within the set $\{ 192,384,768,1536\}$, which does not overlap with the test lengths. The third  approach, which works for fixed input networks,   consists of training a network on the smallest input length, dividing longer signals into chunks, and averaging the predictions of each chunk as proposed  in~\cite{zheng_fusion_2019}.  For that approach, we train DPN from scratch using signals of length 128 and predict exclusively using that length. For signals longer than 128, predictions are applied on chunks of 128 and the results are averaged.
		
		The results for prediction errors and modulation classification are shown in Fig.~\ref{fig:params_pkt}.  For all approaches, as expected the larger the signal length, the lower the estimation error  and the higher the classification accuracy. Looking at the modulation classification results in Fig.~\ref{fig:mod_pkt}, reusing the weights seems to generalize well on lengths larger than 512 without any modification. For shorter lengths, however, the performance drops significantly below 60\%. Retraining the weights on shorter and longer signal lengths, although different from the testing ones, does improve the performance on shorter signals as seen by looking at the retrained curve in the same Figure. This retraining, however, comes  at the cost of  slightly degraded performance on  longer signals.   The relative trend of results between retrained and reused is the same for frequency offset  and symbol duration estimation.  This indicates that DPN  learns signal features that generalize to unseen lengths. By retraining, the weights are adjusted to improve performance on shorter signals at the expense of longer signals.
		
		Now, looking at the curve for dividing the signal into chunks, we see that it only outperforms both the other approaches when $N_r=128$, which is the length it was trained on. Other than that length, it provides higher estimation errors and a lower modulation classification accuracy by up to 15\%.  This shows that this approach is not the best way to use DPN predictions on long signals.  Since the outputs  $\widehat{f}_0$ and $\widehat{\tau}$ are calculated using average pooling, the degradation is not due to the averaging operation itself but to the  weights learned in training. When using only short signals in training, the learned weights extract features leveraging only short signal durations neglecting features spanning  longer periods.  These results highlight the benefit of DPN's design  that can work using different input lengths.

		\section{Conclusion}
		In this paper, we proposed the dual path network (DPN) for blind decoding and modulation classification.  DPN's design  consists of three stages of signal recovery connected to form a signal path made  of DSP operations and a feature path consisting of neural networks. This design enables features to be shared, enabling improved estimates in the later stages and a 5\% improvement in modulation classification compared to a network  separately trained. Extracting features from the recovered signals provides a 5\% improvement in modulation classification compared to a similar network not recovering the signal. The estimation results of DPN  are shown to outperform a reference blind DSP algorithm based on cyclostationary features. These improved estimates make it yield lower packet error rates compared to the reference algorithm.  Due to the  signal processing inspired design,  the output filter taps and estimates are compatible with existing signal processing approaches. Using an over-the-air capture, we validated DPN results at SNRs above 14dB. By relying on average pooling and recurrent neural networks, DPN can process variable length inputs, which are shown to provide better estimates than  dividing the input into short chunks and averaging. 
		
		While this work has considered blind symbol decoding, the dual path concept can be extended to other problems in wireless communications leveraging deep learning. By using a feature path consisting of neural networks and a signal path consisting of DSP operations, the outputs can be made compatible with existing signal processing approaches. Also, this design  inherently enables the reuse of intermediate outputs and feature sharing, which were both shown to create a significant accuracy improvement.
		
		\appendix
		
		\subsection{Band Segmentation}
		\label{ap:band_seg}
		It is applied in two stages using  Welch power spectral density (PSD), which consists of dividing the signal into overlapping segments, calculating the squared magnitude FFT of each segment, and averaging them. In the first stage,  PSD uses an FFT of size $N_1$. The occupied frequency bins are detected using a threshold $T$, with  the edges of the  occupied frequency bins  given by $b_1$ and $b_2$. The center frequency and bandwidth are calculated using  $\frac{b_1 + b_2}{2}$ and $\frac{b_2 - b_1}{2}$, respectively.  The center frequency offset is corrected and a low pass filter is applied to the signal to attenuate the noise. In the second stage, PSD is applied again with a larger FFT of size $N_2>N_1$ to yield a higher resolution frequency and bandwidth estimate using the same procedure. Other than the FFT size, the Welch power spectral used the default parameters in the python SciPy library ~\cite{2020SciPy-NMeth}. We used  $N_1=64$ and  $N_2=256$ and $T=2 N_0$.

		\subsection{Fine Carrier Frequency Estimation}
		\label{ap:freq}
		The estimation of the carrier frequency is performed by detecting cyclostationary features at $\alpha = 4 f_0$~\cite{rebeiz_energy-efficient_2014}. Using the initial frequency estimate, a search window  $\mathcal{W}_{f_0}$ is calculated and the estimated  $4 f_0$ is calculated on the input signal $z[k]$ using
		\begin{equation}
		\underset{\alpha_i\in \mathcal{W}_{f_0}}{\max} \left| \sum_{k=0}^{N_r -1} z[k]^4 e^{-j2\pi\alpha_i k \tau_0} \right|
		\end{equation}
		Given that the coarse frequency estimate from the prior stage was given by $f_{0c}$, the window $\mathcal{W}_{f_0}$ consisted of 100 samples within  $4 f_{0} + \pm0.001$.
		
		\subsection{Fine Symbol Rate Estimation}
		\label{ap:symb_rate}
		The single carrier modulations used in the evaluation have a cylostationnary feature at $\alpha = 1/\tau$, from which $\tau$ can be detected~\cite{rebeiz_energy-efficient_2014}. Using the coarse bandwidth estimate, a search window $\mathcal{W}_{\tau}$ is calculated and the estimated  $1/\tau$ is calculated on the input signal $z[k]$ using
		\begin{equation}
		\underset{\alpha_i\in \mathcal{W}_{\tau}}{\max} \left| \sum_{k=0}^{N_r -1} |z[k]|^2 e^{-j2\pi\alpha_i k \tau_0} \right|
		\end{equation}
		Given that the coarse bandwidth estimate from the prior stage was given by $1/\tau_c$, the window $\mathcal{W}_{\tau}$ consisted of 100 samples between $\frac{0.85}{\tau_c} $ and  $\frac{1.15}{\tau_c} $.

		\subsection{Symbol Timing Offset Estimation}
		\label{ap:timing_off}
		The input signal $z[k]$ is first interpolated by a factor of $P$ to obtain $z_I[k]$. An error is calculated using~\cite{gardner_bpsk/qpsk_1986}
		\begin{equation}
		e[k] =( z_I[k-P/2] -z_I[k+P/2])z_I[k]^*
		\end{equation} 
		where $(\cdot)^*$ denotes the  conjugate. The error vector is divided into windows of size $P$, which are averaged across time. The timing offset index is given by the index of the zero down crossing. We used $P=64$.
		\subsection{Blind Equalization}
		\label{ap:blind_eq}
		The CMA is an iterative  blind equalization algorithm.  At step $m$, it generates $\b{w}(m)$ using   stochastic gradient descent as  follows~\cite{qinghua_shi_blind_2012}
		\begin{equation}
		\b{w}(m) = \b{w}(m-1) - \mu g ( |g| ^2 - 1) \b{r}(m)^*
		\end{equation}
		where $\mu$ is the learning rate,
		$
		g(m) = \b{w}(m-1)^T \b{r}(m)
		$, and
		$\b{r}(m)$ is a slice of  the input signal $z[k]$ starting with index $m$ and of the same length as the filter $\b{w}$. We used $\mu=10^{-4}$ and an equalization filter of length 20. 
		
		\subsection{Genie Equalization}
		\label{ap:genie_eq}
		Let $h[n]$ be the channel taps having Fourier transform $H[k]$. The frequency domain representation of the equalized signal $\widehat{\b{Z}}$ is given by
		$
		\widehat{Z}[k] = \frac{Z[k]H^*[k]}{ H^*[k]H[k]+ N_0 }
		$
		where $\b{Z}$ is the FFT of the input signal. The equalized signal is obtain using the inverse FFT of $\widehat{\b{Z}}$.
		
		\subsection{Symbol Decoding}
		\label{ap:symb_decod}
		Let $\hat{\b{s}}$ denote the recovered symbols before the hard decision, $\b{s}$ the true symbols, $\b{c}_M$ be the constellation of linear modulation $M$.  The filtered phase error at the $k$-th symbol is denoted by  $e_f[k]$. Using the assumed knowledge of the first symbol, we set $e_f[0] = \hat{s}[0] s[0]^* $. We start by calculating the constellation index of the predicted symbol $s_b[k]$ by finding the  minimizer of the Euclidean distance using
		$
		s_b[k] = \text{argmin}  \left| \hat{s}[k] e^{j e_f[k] } - \b{c}_M \right|,
		$
		thus making the decoded symbol
		$
		s_d[k] = c_m [ s_b[k] ]  
		$.
		The phase error $e[k]$ is calculated using the received and the decoded symbol using
		$
		e[k] = \text{arctan} (\hat{s}[k] d[k]^*)
		$, which is then filtered according to
		$
		e_f[k] =   e_f[k] + \alpha e[k]  
		$
		for some constant $\alpha$. The symbol error rate is calculated by comparing the decoded symbols $\b{s}_d$ with the true symbols $\b{s}$. The symbol by symbol comparison is limited to the shortest of both if their lengths are different due to timing errors. We used $\alpha=0.5$.
		
		\vspace{-0.5em}
		\bibliographystyle{IEEEtran}
		\bibliography{references}

	\end{document}